\documentclass{aa}  

\usepackage{graphicx}
\usepackage{txfonts}
\usepackage{hyperref}
\hypersetup{
    colorlinks=true,
    linkcolor=blue,
    filecolor=magenta,      
    urlcolor=cyan,
    citecolor=blue
    }

\usepackage{graphicx}	
\usepackage{amsmath}	
\usepackage[version=4]{mhchem}
\usepackage{gensymb}
\usepackage{chemfig}

\begin{document}

   \title{Beyond monoculture: polydisperse moment methods for sub-stellar atmosphere cloud microphysics}

   \subtitle{I. Examining properties of the exponential distribution}

   \author{Elspeth K.H. Lee\inst{1}}

   \institute{$^{1}$Center for Space and Habitability, University of Bern, Gesellschaftsstrasse 6, CH-3012 Bern, Switzerland}

   \date{Received 13 March 2025 / Accepted 07 May 2025}

 
  \abstract
   {Observational data provided by JWST instruments continue to challenge theories and models of cloud formation in sub-stellar atmospheres, requiring more sophisticated approaches in an effort to understand their spatial complexity.
   However, to date, most cloud microphysical models using the moment method for sub-stellar atmospheres have assumed a monodisperse size distribution, neglecting polydisperse properties.}
   {We aim to extend beyond the common assumption of a monodisperse size distribution and analyse cloud microphysical processes assuming an exponential distribution.}
   {We derive expressions for the zeroth and first moments of condensation/evaporation and collisional growth processes under the assumption of an exponential size distribution.
   We then compare the differences between monodisperse and exponential distribution microphysics using a simple one-dimensional (1D) column model applied to a Y-dwarf KCl cloud scenario.}
   {We find that adopting an exponential distribution modifies condensation/evaporation rates by a factor of $\approx$0.9 and collisional growth rates by factors of $>$1.1 (Kn $\ll$ 1) and $\approx$1.37 (Kn $\gg$ 1) for Brownian coagulation and $\approx$0.85 for gravitational coalescence, compared to the monodisperse case.
   In our specific test cases, we find maximal relative differences of $>$200\% in total number density and $>$40\% in mean radius of the cloud particles between the monodisperse and exponential distributions.}
   {Our framework offer a simple way to take into account polydispersity with an assumed exponential size distribution for sub-stellar atmospheric cloud microphysics using a two-moment method.
   In follow up studies, we will examine more complex distributions, such as the log-normal and gamma distributions, that require more than two moments to characterise self-consistently.}


   \authorrunning{Elspeth K.H. Lee}
   \maketitle
%

\section{Introduction}

The cloud particle size distribution plays a pivotal role in setting the observational properties of atmospheres that contain a cloud component, as the total opacity of the cloud is weighted by the individual contribution of each particle size to the size distribution. 
Further knock-on effects, affecting the thermal structure of the atmosphere through feedback with the radiative-transfer properties of the atmospheres, which, for example, in the case for brown dwarf atmospheres, can trigger local convection \citep[e.g.][]{Tan2019} leading to changes in local chemical compositions \citep[e.g.][]{Lee2024}.
This makes the consideration and accurate modelling of cloud properties a key component in contemporary exoplanet and brown dwarf atmospheric modelling.

Exponential size distributions have been used for modelling Earth-based clouds, particularly raindrops, for decades since \citet{Marshall1948} proposed an exponential distribution as a best fit to observed raindrop data.
Exponential distributions have consistently been applied to model raindrops and snow in the Earth atmosphere literature \citep[e.g.][]{Pruppacher1978,Lin1983,Hong2006,Seifert2008,Morrison2020}, with these schemes being used in general circulation models (GCMs) and other atmosphere simulation efforts such as mesoscale and large eddy simulations.
However, for exoplanet and brown dwarf atmospheres the exact particle size distribution is unknown and is currently not directly measurable with sufficient accuracy.
Therefore, instead of parametrisation and fitting from in-situ data, as typically performed for Earth water clouds, it is advisable to develop theories and models in a generally applicable manner and self-consistent within their specific assumptions.

The moment (or bulk) method evolves integrated quantities of the particle size distribution over time, considering how cloud microphysical processes such as nucleation, condensation, evaporation, and collisions affect these quantities.
The moment method is in contrast to the bin/spectral method which directly integrates a discretised grid of particle sizes or masses.
A well used example of a bin model for sub-stellar objects is the 1D CARMA model \citep[e.g.][]{Gao2018, Powell2019}. 
However, the bin method is typically too computationally expensive to couple to large scale hydrodynamic models such as GCMs with contemporary resources, making development of moment schemes a useful middle-ground between computational efficiency and understanding the spacial time evolution of cloud microphysics.

Examples of moment methods used in sub-stellar atmosphere literature context ranges from 1D static approaches \citep[e.g.][]{Woitke2004, Helling2008}, 1D time-dependent approaches \citep[e.g.][]{Ohno2017, Woitke2020} to 3D time-dependent methods \citep[e.g.][]{Lee2016, Lines2018, Lee2023}.
Several studies on sub-stellar atmospheres have employed multiple moment methods derived for monodisperse size distributions and subsequently reconstructed the particle size distribution using the moment solutions \citep[e.g.][]{Helling2008, Stark2015, Lee2023}.

Previous efforts typically assumed a delta function for a monodisperse size distribution to derive a mass moment equation set for the cloud microphysical system.
In this short study, we go beyond a monodisperse size distribution and explore how the assumption of an exponential particle size distribution affects the condensation/evaporation and collisional growth rates of cloud particles for the mass moment method.
In Section \ref{sec:exp_dist}, we introduce the exponential distribution and its useful properties when applied to the moment method framework.
In Section \ref{sec:exp_cond}, we derive modifications to the monodisperse size distribution mass condensation/evaporation rates for an exponential distribution. 
We also propose a tanh function that smoothly interpolates between the continuum/diffusive and free molecular/kinetic regime limits near the critical Knudsen number.
In Section \ref{sec:exp_coll}, we derive expressions for collisional growth for Brownian coagulation and gravitational coalescence.
In Section \ref{sec:set_int}, we propose an interpolation function for the settling velocity between the continuum/diffusive and free molecular/kinetic regime limits.
In Section, \ref{sec:1D}, we apply our new theory to simple 1D models for a Y-dwarf KCl cloud formation case.
Lastly, Sections \ref{sec:disc} and \ref{sec:conc} respectively contain the discussion and conclusion of our study.
The modelling and code from this paper can be found as part of the `mini-cloud' framework in the GitHub repository\footnote{{\url{https://github.com/ELeeAstro/mini_cloud}}}.

\section{Moments with assumed exponential distribution}
\label{sec:exp_dist}

\begin{figure*}
    \centering
    \includegraphics[width=0.49\linewidth]{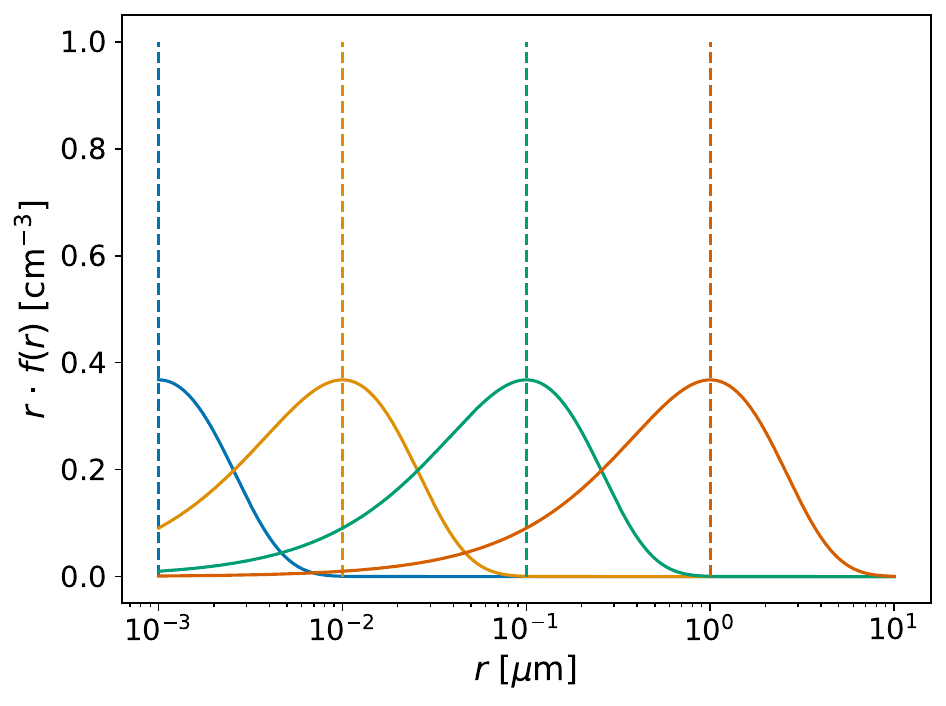}
    \includegraphics[width=0.49\linewidth]{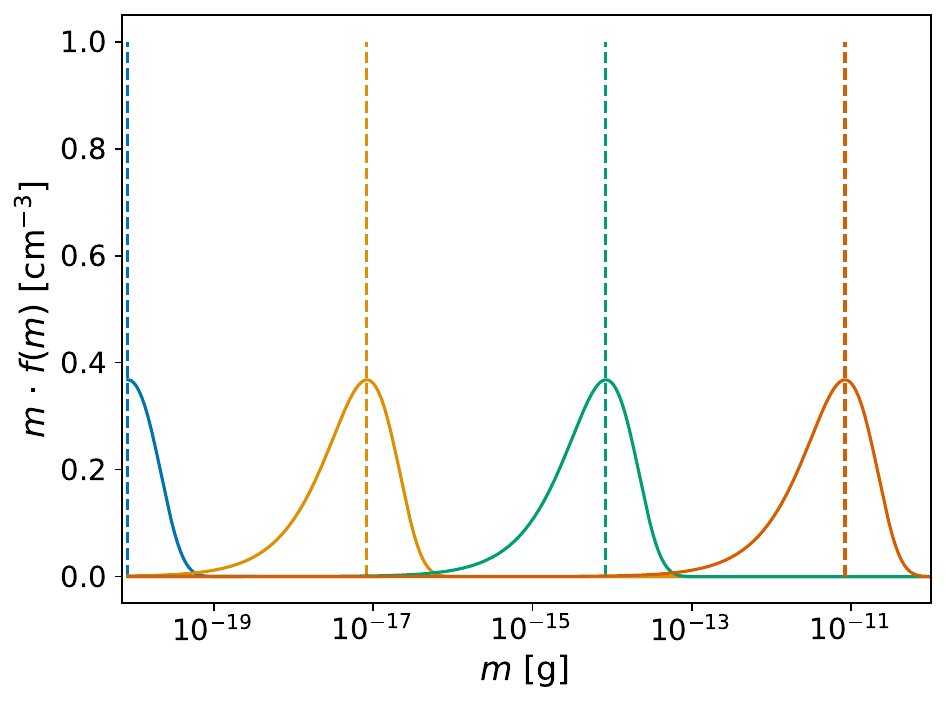}
    \caption{Visualisation of the exponential size distribution compared against a monodisperse size distribution in size units (left) and mass units (right). 
    This shows exponential distributions with (solid lines) $\lambda$ = 10$^{-3}$ (blue), 10$^{-2}$ (orange), 0.1 (green) and 1 (red) $\mu$m assuming $N_{\rm c}$ = 1 cm$^{-3}$ and $\rho_{\rm d}$ = 1 g cm$^{-3}$, with the equivalent monodisperse distribution (dashed lines).
    Overall, the exponential distribution enhances the representation of smaller particle sizes relative to the monodisperse distribution. 
    The particle mass distribution is spread over more orders of magnitude than the particle size distribution.}
    \label{fig:exp_mono}
\end{figure*}

In this Section, we present a derivation of the mass moments assuming an exponential distribution, as well as derive the source terms required for evolving the cloud formation microphysics ordinary differential equation (ODE) system in time.
The exponential distribution, $f(m)$ [cm$^{-3}$ g$^{-1}$], is a special single parameter form of the gamma distribution given by
\begin{equation}
\label{eq:exp_dist}
    f(m) = \frac{N_{\rm c}}{\lambda}e^{-m/\lambda},
\end{equation}
where $N_{\rm c}$ [cm$^{-3}$] is the total number density of the distribution, $\lambda$ [g] is the mean particle mass, which is equal to the expectation value, E[m], of the distribution, and m [g], an individual mass in the distribution.
The exponential distribution is a favourable distribution when attempting to derive a set of moment equations for a cloud microphysics system due its single parameter and useful analytical properties.

The mass moments of the particle mass distribution, $M^{(k)}$ [g$^{k}$ cm$^{-3}$], are given by
\begin{equation}
    M^{(k)} = \int_{0}^{\infty}m^{k}f(m)dm,
\end{equation}
where $k$ is the moment power.
Inserting the exponential distribution in the moment equation gives
\begin{equation}
    M^{(k)} = \frac{N_{\rm c}}{\lambda}\int_{0}^{\infty}m^{k}e^{-m/\lambda}dm.
\end{equation}
After some algebra and applying the definition of the gamma integral\footnote{$\Gamma(z)$ = $\int_{0}^{\infty}t^{z-1}e^{-t}dt$}, the moment generator for the exponential size distribution is analytically expressed as
\begin{equation}
\label{eq:mom_gen}
    M^{(k)} = N_{\rm c}\lambda^{k}\Gamma(k + 1),
\end{equation}
where $\Gamma$ is the gamma function.
The moment generator is useful to close the moment system when required through estimating the higher moment values from lower ones, for example, when calculating the moment dependent settling velocity \citep[e.g.][]{Woitke2020}.
Each integer moment represents bulk values of the particle size distribution, such as the total number density, $N_{\rm c}$ [cm$^{-3}$], for $k$ = 0
\begin{equation}
    M^{(0)} = N_{\rm c},
\end{equation}
and the $k$ = 1 moment which represents the total mass density, $\rho_{\rm c}$ [g cm$^{-3}$], of the cloud particles
\begin{equation}
  M^{(1)} = \rho_{\rm c}.
\end{equation}

The value of $\lambda$ can be readily estimated from the moment solutions as 
\begin{equation}
    \lambda = \frac{M^{(1)}}{M^{(0)}} = m_{\rm c},
\end{equation}
which closes the system of equations. 
Here, following the notation of \citet{Lee2025}, $m_{\rm c}$ [g] is the average mass of the particles.
The mean mass-weighted average particle size, $r_{\rm c}$ [cm], is then
\begin{equation}
\label{eq:rc}
    r_{\rm c} = \left(\frac{3m_{\rm c}}{4\pi\rho_{\rm d}}\right)^{1/3},
\end{equation}
where $\rho_{\rm d}$ [g cm$^{-3}$] is the particle bulk density.
Throughout this study, we assume spherical particles of a homogeneous composition using Eq. \ref{eq:rc}, we do not consider non-spherical or porous particles.

In Fig. \ref{fig:exp_mono}, we show visually the difference between the monodisperse and exponential distributions.
From this, it is clear that the monodisperse distribution through its assumption of a single representative grain size places all emphasis on the mean particle size, whereas the exponential distribution is more `realistic' in that there is a variance around the mean particle size.
In addition, from Fig. \ref{fig:exp_mono} the particle mass distribution is generally spread across a large range of values compared to the particle size distribution.
In the exponential distribution, this variance is constant and simply given by the square of the average value
\begin{equation}
    {\rm Var}[m] = \lambda^{2}.
\end{equation}
This single parameter property makes the exponential distribution useful for two moment methods where only the mean particle mass/size can be estimated from the moment solutions.
The exponential distribution is characterised by a gradual increase up to the mean particle size, after which the particle number density drops quickly. 
This places greater emphasis on the small particle population than the monodisperse distribution, a factor that must be considered when modelling physical processes affecting the size distribution.
However, this single parameter approach has significant limitations on the flexibility of the size distribution (Section \ref{sec:disc}).

\section{Condensation with exponential distribution}
\label{sec:exp_cond}

In this Section, we derive the rate of condensation/evaporation of mass from the cloud bulk when assuming an exponential distribution and compare it to the monodisperse derivation.
Following the differentiation with time for the method of moments \citep[e.g.][]{Gail2013}, the change in time of the moments for the condensation/evaporation process is given by
\begin{equation}
\label{eq:dMk_dt}
    \frac{dM^{(k)}}{dt} = k\int_{0}^{\infty}m^{k-1}\frac{dm}{dt}f(m)dm,
\end{equation}
where, $dm/dt$ [g s$^{-1}$], is the rate of change of the cloud particle mass.
In the monodisperse size distribution, the particle size distribution is assumed to be a delta function at the mean particle size expressed formally as 
\begin{equation}
    f(m) = N_{\rm c}\delta(m - \lambda).
\end{equation}
$dm/dt$ is then independent of mass, removing it from the integral in Eq. \ref{eq:dMk_dt}.
This leads to the monodisperse moment method typically used in the literature \citep[e.g.][]{Woitke2003,Helling2008,Gail2013, Ohno2017, Lee2023}, where the rate of change of a moment $k$ is
\begin{equation}
\label{eq:dMdt_mono}
    \frac{dM^{(k)}}{dt} = k\frac{dm}{dt}\int_{0}^{\infty}m^{k-1}f(m)dm = k\frac{dm}{dt}M^{(k-1)}.
\end{equation}
For example, the monodisperse size distribution source term for condensational growth/evaporation for the 1st moment ($k$ = 1) is
\begin{equation}
\label{eq:Kns_cond_mono}
    \frac{d\rho_{\rm c}}{dt} =  \left.\frac{dm}{dt}\right|_{r_{\rm c}}N_{\rm c},
\end{equation}
where $dm/dt$ is given by Eq. \ref{eq:cond_d} or \ref{eq:cond_f} dependent on the Knudsen number regime and evaluated at the mean particle size $r_{\rm c}$.

We define the Knudsen number, Kn, as 
\begin{equation}
    {\rm Kn} = \frac{\lambda_{\rm a}}{r},
\end{equation}
where $r$ [cm] is the particle radius, and $\lambda_{\rm a}$ [cm] is the atmospheric mean path length which can be given by
\begin{equation}
    \lambda_{\rm a} = \frac{2\eta_{\rm a}}{\rho_{\rm a}}\sqrt{\frac{\pi\bar{\mu}_{\rm a}}{8RT}},
\end{equation}
with $\eta_{\rm a}$ [g cm$^{-1}$ s$^{-1}$] the atmospheric dynamical viscosity which be calculated following \citet{Lee2025} using the \citet{Rosner2012} fitting function and \citet{Davidson1993} mixing rule, $\rho_{\rm a}$ [g cm$^{-3}$] the atmospheric mass density and $\bar{\mu}_{\rm a}$ [g mol$^{-1}$] the mean molecular weight of the atmosphere.

The monodisperse Knudsen number is given by
\begin{equation}
    \overline{{\rm Kn}} = \frac{\lambda_{\rm a}}{r_{\rm c}},
\end{equation}
and we can define a population number averaged Knudsen number, $\overline{{\rm Kn}_{N}}$, as
\begin{equation}
    \overline{{\rm Kn}_{N}} = \frac{\int_{0}^{\infty}{\rm Kn}(m)f(m)dm}{\int_{0}^{\infty}f(m)dm} = \frac{\lambda_{\rm a}s\int_{0}^{\infty}m^{-1/3}f(m)dm}{N_{\rm c}},
\end{equation}
where $s$ = $(3/4\pi\rho_{\rm d})^{-1/3}$.
Using the definition of the moments and the Eq. \ref{eq:mom_gen} moment generator, this results in the expression
\begin{equation}
    \overline{{\rm Kn}_{N}} = \frac{\lambda_{\rm a}sM^{(-1/3)}}{N_{\rm c}} = \frac{\lambda_{\rm a}}{r_{\rm c}}\Gamma\left(\frac{2}{3}\right) = \overline{{\rm Kn}}\Gamma\left(\frac{2}{3}\right) ,
\end{equation}
where $M^{(-1/3)}$ is the $k$ = -1/3 non-integer moment power.
Similarly, the mass-weighted Knudsen number, $\overline{{\rm Kn}_{m}}$,  is given by
\begin{equation}
\label{eq:Kn_m1}
    \overline{{\rm Kn}_{m}} = \frac{\int_{0}^{\infty}{\rm Kn}(m)mf(m)dm}{\int_{0}^{\infty}mf(m)dm} = \frac{\lambda_{\rm a}s\int_{0}^{\infty}m^{2/3}f(m)dm}{\rho_{\rm c}},
\end{equation}
with
\begin{equation}
\label{eq:Kn_m2}
    \overline{{\rm Kn}_{m}} = \frac{\lambda_{\rm a}sM^{(2/3)}}{\rho_{\rm c}} = \frac{\lambda_{\rm a}}{r_{\rm c}}\Gamma\left(\frac{5}{3}\right) = \overline{{\rm Kn}}\Gamma\left(\frac{5}{3}\right),
\end{equation}
where $M^{(2/3)}$ is the $k$ = 2/3 non-integer moment power.

\subsection{Kn $\ll$ 1 regime}

For the exponential distribution (and others), unlike for the monodisperse distribution, the mass dependence in the growth rate cannot be ignored in the integral of Eq. \ref{eq:dMk_dt}.
The change in mass for condensation/evaporation in the continuum/diffusive regime (Kn $\ll$ 1) is given by \citep[e.g.][]{Woitke2003}
\begin{equation}
\label{eq:cond_d}
    \left(\frac{dm}{dt}\right)^{\rm diff} = 4\pi r D m_{\rm v} n_{\rm v} \left(1 - \frac{1}{S}\right),
\end{equation}
where $r$ [cm] is the radius of the particle, $m_{\rm v}$ [g] the mass of one unit of vapour, $n_{\rm v}$ [cm$^{-3}$] the number density of the vapour and $S$ the supersaturation ratio of the vapour.
$D$ [cm$^{2}$ s$^{-1}$] is the diffusive rate of the vapour onto the particle surface and can be given by \citep[e.g.][]{Jacobson2005}
\begin{equation}
    D = \frac{5}{16N_{\rm A}d_{i}^{2}\rho_{\rm a}}\sqrt{\frac{RT\bar{\mu}_{\rm a}}{2\pi}\left(\frac{\mu_{\rm v} + \bar{\mu}_{\rm a}}{\mu_{\rm v}}\right)},
\end{equation}
where $N_{\rm A}$ is Avogadro's number, $d_{i}$ [cm] the collision diameter, and $\mu_{v}$ [g mol$^{-1}$] the molecular weight of the condensable vapour.

The particle radius is converted to particle mass through the relation
\begin{equation}
    r = \left(\frac{3m}{4\pi\rho_{\rm d}}\right)^{1/3},
\end{equation}
which reveals the mass dependence, $\propto$ $m^{1/3}$, of condensational growth/evaporation in the diffusive/continuum regime.
Defining the mass independent kinetic pre-factor as $C_{0}$
\begin{equation}
    C_{0} = 4\pi \left(\frac{3}{4\pi\rho_{\rm d}}\right)^{1/3} D m_{\rm v} n_{\rm v} \left(1 - \frac{1}{S}\right),
\end{equation}
Eq. \ref{eq:dMk_dt} becomes
\begin{equation}
  \frac{dM^{(k)}}{dt} = kC_{0}\int_{0}^{\infty}m^{1/3}m^{k-1}f(m)dm = kC_{0}M^{(k-2/3)}.
\end{equation}

\subsubsection{Monodisperse solution}
For the monodisperse assumption, Eq. \ref{eq:dMdt_mono} is used. This leads to the change in the first moment ($k$ = 1) being
\begin{equation}
\label{eq:qcdt_l}
    \left(\frac{d\rho_{\rm c}}{dt}\right)_{\rm cond}^{\rm diff} = 4\pi r_{\rm c} D m_{\rm v} n_{\rm v} \left(1 - \frac{1}{S}\right)N_{\rm c}.
\end{equation}

\subsubsection{Exponential distribution solution}
Using the exponential distribution moment generator (Eq. \ref{eq:mom_gen}), Eq. \ref{eq:dMk_dt} is now
\begin{equation}
\label{eq:diff_mono}
    \frac{dM^{(k)}}{dt} = kC_{0}\lambda^{k-2/3}N_{\rm c}\Gamma(k + 1/3).
\end{equation}
We can then use Eq. \ref{eq:mom_gen} for the $k-1$ moment to factor out $N_{\rm c}$ and get the expression in terms of the previous moment.
\begin{equation}
    \frac{dM^{(k)}}{dt} = kC_{0}\lambda^{1/3}M^{(k-1)}\frac{\Gamma(k + 1/3)}{\Gamma(k)}.
\end{equation}
The full expression for the first moment ($k$ = 1) is then
\begin{equation}
\label{eq:qcdt_l}
    \left(\frac{d\rho_{\rm c}}{dt}\right)_{\rm cond}^{\rm diff} = 4\pi r_{\rm c} D m_{\rm v} n_{\rm v} \left(1 - \frac{1}{S}\right)\Gamma\left(\frac{4}{3}\right)N_{\rm c},
\end{equation}
which reveals that the rate of change of the first moment in the Kn $\ll$ 1 regime for the exponential distribution is a factor of $\Gamma(4/3)$ different to the monodisperse rate (Eq. \ref{eq:diff_mono}).

\subsection{Kn $\gg$ 1 regime}
In the large Knudsen number regime, Kn $\gg$ 1, the free molecular/kinetic regime, the mass dependence is $\propto$ $m^{2/3}$ as the condensation rate expression is given by \citep[e.g.][]{Woitke2003}
\begin{equation}
\label{eq:cond_f}
    \left(\frac{dm}{dt}\right)^{\rm free} = 4\pi r^{2}v_{\rm th}m_{\rm v} n_{\rm v}\alpha \left(1 - \frac{1}{S}\right),
\end{equation}
where $v_{\rm th}$ = $\sqrt{k_{\rm b}T/2\pi m_{v}}$ [cm s$^{-1}$] is the thermal velocity of the condensable vapour, with $m_{v}$ [g] the mass of the condensable vapour.
$\alpha$ = [0,1] is the `sticking efficiency' of the vapour to the particle surface.
Again, defining the constant kinetic pre-factor $C_{0}$ as
\begin{equation}
    C_{0} = 4\pi \left(\frac{3}{4\pi\rho_{\rm d}}\right)^{2/3}v_{\rm th}m_{\rm v} n_{\rm v} \alpha \left(1 - \frac{1}{S}\right),
\end{equation}
Eq. \ref{eq:dMk_dt} becomes
\begin{equation}
  \frac{dM^{(k)}}{dt} = kC_{0}\int_{0}^{\infty}m^{2/3}m^{k-1}f(m)dm = kC_{0}M^{(k-1/3)}.
\end{equation}

\subsubsection{Monodisperse solution}
As in the Kn $\ll$ 1 regime, \ref{eq:dMdt_mono} is used for the monodisperse distribution.
This leads to the change in the first moment ($k$ = 1) being
\begin{equation}
    \left(\frac{d\rho_{\rm c}}{dt}\right)_{\rm cond}^{\rm free} = 4\pi r_{\rm c}^{2} v_{\rm th}m_{\rm v} n_{\rm v} \left(1 - \frac{1}{S}\right)N_{\rm c}.
\end{equation}

\subsubsection{Exponential distribution solution}

Similar to the Kn $\ll$ 1 regime solution, after using the moment generator equation (Eq. \ref{eq:mom_gen}) the moment expression is given as
\begin{equation}
    \frac{dM^{(k)}}{dt} = kC_{0}\lambda^{k-1/3}N_{\rm c}\Gamma(k + 2/3),
\end{equation}
and in terms of the $k-1$ moment
\begin{equation}
    \frac{dM^{(k)}}{dt} = kC_{0}\lambda^{2/3}M^{(k-1)}\frac{\Gamma(k + 2/3)}{\Gamma(k)}.
\end{equation}
This gives for the full first moment ($k$ = 1) expression
\begin{equation}
\label{eq:qcdt_h}
    \left(\frac{d\rho_{\rm c}}{dt}\right)_{\rm cond}^{\rm free} = 4\pi r_{\rm c}^{2} v_{\rm th}m_{\rm v} n_{\rm v} \left(1 - \frac{1}{S}\right)\Gamma\left(\frac{5}{3}\right)N_{\rm c},
\end{equation}
which reveals that the first moment in the Kn $\gg$ 1 regime is a factor of $\Gamma(5/3)$ different when assuming an exponential size distribution compared to the monodisperse distribution.

Overall, the difference between the monodisperse rate of change and exponential assumed distribution for the first moment is the factors of gamma, $\Gamma(4/3)$ and $\Gamma(5/3)$, which are both $<$ 1. 
The rate of condensation in the exponential distribution is therefore a factor $\sim$0.9 slower than the monodisperse distribution.
This makes intuitive sense, as the monodisperse distribution will weight the number of particles at the mean value compared to the exponential distribution which weights towards more smaller particle representation in its formulation.
Combined with the exponential distributions strong drop-off in particle size after the mean value, the condensation rate for the exponential distribution is expected to be lower than the monodisperse distribution.

Our derived analytical solution makes it trivial to include an exponential particle size distribution dependence in current exoplanet atmosphere cloud condensation codes that use the moment method.
Analogous derivations in particle radius or volume space instead of mass are a possibility for theories and models that apply the moments in radius or volume space \citep[e.g.][]{Helling2008,Gail2013}.

\subsection{Condensation regime interpolation function}

Equations \ref{eq:qcdt_l} and \ref{eq:qcdt_h} are valid both in their respective Knudsen number regimes of Kn $\ll$ 1 and Kn $\gg$ 1 respectively. 
However, in the intermediate regime, Kn $\sim$ 1, a direct expression for the mass condensation/evaporation rate equation is difficult to derive.

To address this, following \citet{Woitke2003}, we apply an interpolation function that transitions smoothly between the diffusive and free molecular regimes.
When the expressions Eq. \ref{eq:qcdt_l} and Eq. \ref{eq:qcdt_h} are equal, this results in \citep{Woitke2003}
\begin{equation}
    \alpha v_{\rm th} r = D.
\end{equation}
Using the Knudsen number definition, the critical Knudsen number, ${\rm Kn_{\rm cr}}$, for the monodisperse distribution ($r$ = $r_{\rm c}$) is given by
\begin{equation}
    {\rm Kn_{\rm cr}} = \frac{\lambda_{\rm a}\alpha v_{\rm th}}{D},
\end{equation}
and for the population averaged exponential distribution is
\begin{equation}
    \overline{{\rm Kn_{\rm cr}}} = \frac{\lambda_{\rm a}\alpha v_{\rm th}\Gamma\left(5/3\right)}{D\Gamma\left(4/3\right)}.
\end{equation}

The critical Knudsen number can then be used to interpolate between each regime near the critical value \citep[e.g.][]{Woitke2003}.
We propose a symmetric smooth tanh function of the form
\begin{equation}
\label{eq:tanh}
    f({\rm Kn}') = \frac{1}{2}[1 - \tanh(a\log_{\rm 10}\rm{Kn}')],
\end{equation}
where $\rm{Kn}'$ = $\overline{{\rm Kn}}$/Kn$_{\rm cr}$ for the monodisperse formulation, or $\overline{\rm{Kn}'}$ = $\overline{{\rm Kn}_{m}}$/$\overline{{\rm Kn}_{\rm cr}}$ for the population averaged value, and has a range between 0 and 1, with $\overline{{\rm Kn}_{m}}$ the mass-weighted population averaged Knudsen number (Eq. \ref{eq:Kn_m2}).
For the monodisperse distribution, the net change in the first moment is then a linear combination using this interpolation function
\begin{multline}
\label{eq:2nd_int}
    \left(\frac{d\rho_{\rm c}}{dt}\right)_{\rm cond} = f({\rm Kn}')\left.\frac{dm}{dt}\right|_{r_{\rm c}}^{\rm diff}N_{\rm c}  \\ + \left[1 - f({\rm Kn}')\right] \left.\frac{dm}{dt}\right|_{r_{\rm c}}^{\rm free}N_{\rm c}.
\end{multline}

We set a scaling value of $a$ = 2, which smoothly transitions between the asymptotic expressions at around 2 orders magnitude in Knudsen number away from the critical number.
At 1 order of magnitude away from the critical number, the interpolation function gives a value of $f({\rm Kn}')$ $\approx$ 0.1 or 0.9.
For consideration of the exponential distribution, the respective gamma factors for each regime can be easily re-included into Eq. \ref{eq:2nd_int}
\begin{multline}
\label{eq:2nd_int2}
    \left(\frac{d\rho_{\rm c}}{dt}\right)_{\rm cond} = f(\overline{{\rm Kn}'})\left.\frac{dm}{dt}\right|_{r_{\rm c}}^{\rm diff}\Gamma\left(\frac{4}{3}\right)N_{\rm c}  \\ + \left[1 - f(\overline{{\rm Kn}'})\right] \left.\frac{dm}{dt}\right|_{r_{\rm c}}^{\rm free}\Gamma\left(\frac{5}{3}\right)N_{\rm c}.
\end{multline}

\section{Collisional growth with exponential distribution}
\label{sec:exp_coll}

Another important process in shaping the particle size distribution is the collision and subsequent growth of particles, either through Brownian motion of particles (coagulation) or relative settling velocities of different sized particles (coalescence).
In \citet{Drake1972}, the mass moment generator expression of the Smoluchowski equation for collisional growth is derived as
\begin{equation}\label{eq:coll_mom}
\begin{split}
    \frac{dM^{(k)}}{dt} = &\frac{1}{2}\int_{0}^{\infty}\int_{0}^{\infty}K(m,m')f(m)f(m') \\ & \times \left[(m + m')^{k} - m^{k} - m'^{k}\right]dmdm',
\end{split}
\end{equation}
where $m$ and $m'$ represent different masses in the particle mass distribution and $K(m,m')$ [cm$^{3}$ s$^{-1}$] the kernel function.
We can immediately find the expression for the zeroth moment as
\begin{equation}
\label{eq:N_coll}
    \frac{dM^{(0)}}{dt} = -\frac{1}{2}\int_{0}^{\infty}\int_{0}^{\infty}K(m,m')f(m)f(m')dmdm',
\end{equation}
which in general serves as the basis for deriving the collisional growth rate expressions using moment methods.

In \citet{Lee2025}, following arguments in \citet{Rossow1978}, it was assumed that the size distribution is monodisperse, allowing the kernel function to be moved outside the integral.
In this study, we apply the method found in \citet[][Appendix C]{Moran2023} to find size distribution averaged collision kernels, which we repeat below for completeness.
Following Eq. \ref{eq:N_coll}, in \citet{Moran2023} the number density population averaged collision kernel is given as
\begin{multline}
\label{eq:pop_av_k}
       \overline{K(m,m')} = \frac{\int_{0}^{\infty}\int_{0}^{\infty}K(m,m')f(m)f(m')dmdm'}{\int_{0}^{\infty}\int_{0}^{\infty}f(m)f(m')dmdm'} \\
    =  \frac{1}{N_{\rm c}^{2}}\int_{0}^{\infty}\int_{0}^{\infty}K(m,m')f(m)f(m')dmdm'.
\end{multline}

In the following Sections, we use the \citet{Moran2023} technique to derive population average collision kernels for both the Kn $\gg$ 1 and Kn $\ll$ 1 regimes. 
We then find collisional growth rate expressions for the monodisperse and exponential distribution cases.

\subsection{Kn $\ll$ 1 regime}

For collisional growth driven by Brownian motion (coagulation) in the Kn $\ll$ 1 regime, the collisional kernel is given by \citep{Chandrasekhar1943}
\begin{equation}
\label{eq:Diff_k}
    K(r,r') = 4\pi\left[D(r)  + D(r')\right](r + r'),
\end{equation}
where $r$ and $r'$ represent different particle radii in the distribution and $D(r)$ is the particle diffusion factor given by \citep{Chandrasekhar1943}
\begin{equation}
    D(r) = \frac{k_{\rm b}T\beta}{6\pi\eta_{\rm a}r},
\end{equation}
where $\beta$, the Cunningham slip factor, generally depends on the radius (and therefore mass) through scaling with Knudsen number in the functional form
\begin{equation}
    \beta = 1 + {\rm Kn}[A + Be^{-C/{\rm Kn}}].
\end{equation}
In this study, we use the parameters experimentally measured between Kn $\approx$ 0.5-83 from \citet{Kim2005} 
\begin{equation}
\label{eq:C_slip1}
    \beta = 1 + {\rm Kn}[1.165 + 0.483e^{-0.997/{\rm Kn}}],
\end{equation}
which is approximated to within $\approx$10\% in the linear form
\begin{equation}
\label{eq:C_slip2}
    \beta = 1 + A{\rm Kn},
\end{equation}
where A = 1.639, calculated from an optimised least-squares fitting to the \citet{Kim2005} coefficients.

Converting Eq. \ref{eq:Diff_k}  to mass units gives the mass dependence of the kernel
\begin{equation}
    K(m,m') = K_{0}(\beta m^{-1/3} + \beta'm'^{-1/3})(m^{1/3} + m'^{1/3}),
\end{equation}
where $K_{0}$ is the constant kinetic pre-factor given by
\begin{equation}
    K_{0} = \frac{2k_{\rm b}T}{3\eta_{\rm a}}.
\end{equation}

\subsubsection{Monodisperse distribution solution}

If we make the monodisperse assumption for the particle mass distribution, where $f(m)$ $\approx$ $N_{\rm c}$$\delta(m - \lambda)$ and $f(m')$ $\approx$ $N_{\rm c}$$\delta(m' - \lambda)$, the kernel function becomes
\begin{equation}
    K(m,m') = 4K_{0}\overline{\beta},
\end{equation}
where $\overline{\beta}$ is Eq. \ref{eq:C_slip1} evaluated at the mean particle size, leading to a zeroth moment expression given by
\begin{equation}
    \frac{dM^{(0)}}{dt} = - 2K_{0}\overline{\beta}N_{\rm c}^{2}.
\end{equation}
This results in the same equation from \citet{Lee2025} for the number density collisional growth rate
\begin{equation}
\label{eq:N_coag1}
    \left(\frac{dN_{\rm c}}{dt}\right)_{\rm coag} \approx - \frac{4k_{\rm b}T\overline{\beta}}{3\eta_{\rm a}}N_{\rm c}^{2},
\end{equation}
 recovering the monodisperse limit.

\subsubsection{Exponential distribution solution}

We now attempt a zeroth moment solution assuming an exponential size distribution.
Following \citet{Moran2023}, the population averaged kernel in this regime is given by
\begin{multline}
      \overline{K(m,m')} = \frac{1}{N_{\rm c}^{2}}\int_{0}^{\infty}\int_{0}^{\infty}K(m,m')f(m)f(m')dmdm' \\ 
      = \frac{K_{0}}{N_{\rm c}^{2}}\int_{0}^{\infty}\int_{0}^{\infty}(\beta m^{-1/3} + \beta'm'^{-1/3})(m^{1/3} + m'^{1/3}) \\ \cdot f(m)f(m')dmdm'.
\end{multline}
Assuming $\beta$ and $\beta'$ are well represented at the mean particle size, this can be converted into a function of non-integer moment powers as \citep{Moran2023}
\begin{equation}
    \overline{K(m,m')} = \frac{2K_{0}\overline{\beta}}{N_{\rm c}^{2}}\left(M^{(0)}M^{(0)} + M^{(1/3)}M^{(-1/3)}\right),
\end{equation}
where $M^{(1/3)}$ and $M^{(-1/3)}$ are the $k$ = 1/3 and $k$ = -1/3 non-integer power moment respectively.
Using the moment generator for the exponential mass distribution (Eq. \ref{eq:mom_gen}), we can immediately find the population averaged kernel for the exponential distribution in the Kn $\ll$ 1 regime
\begin{equation}
\label{eq:K_av_h}
    \overline{K(m,m')} = 2K_{0}\overline{\beta}\left[1 + \Gamma\left(\frac{4}{3}\right)\Gamma\left(\frac{2}{3}\right)\right].
\end{equation}
Putting this averaged kernel into Eq. \ref{eq:N_coll} gives the change in the number density as
\begin{equation}
\label{eq:N_coag_exp_1}
    \left(\frac{dN_{\rm c}}{dt}\right)_{\rm coag} \approx - \frac{2k_{\rm b}T\overline{\beta}}{3\eta_{\rm a}}\left[1 + \Gamma\left(\frac{4}{3}\right)\Gamma\left(\frac{2}{3}\right)\right]N_{\rm c}^{2}.
\end{equation}
Evaluating the bracket containing the gamma functions gives $\approx$2.21, meaning the rate of change of number density from collisional growth assuming an exponential distribution is therefore around
$\approx$1.1 times larger than the monodisperse assumption in the Kn $\ll 1$ regime.

\subsubsection{Accounting for the Cunningham slip factor}

Assuming the Cunningham slip factor in its linear form (Eq. \ref{eq:C_slip2}) enables a moment analytical solution including the effects of slip correction.
Integrating including the $\beta$ and $\beta'$ factors gives the population averaged kernel as
\begin{multline}
\label{eq:K_av_h2}
    \overline{K(m,m')} = \frac{2K_{0}}{N_{\rm c}^{2}}\Biggl[M^{(0)}M^{(0)} + M^{(1/3)}M^{(-1/3)} \\ + \lambda_{\rm a}A\left(\frac{3}{4\pi\rho_{\rm d}}\right)^{-1/3}\left(M^{(0)}M^{(-1/3)} + M^{(1/3)}M^{(-2/3)}\right)\Biggr],
\end{multline}
which is Eq. \ref{eq:K_av_h} but with an additional moment dependent correction term.
The change in number density with time is then
\begin{multline}
\label{eq:N_coag_exp_2}
    \left(\frac{dN_{\rm c}}{dt}\right)_{\rm coag} \approx - \frac{2k_{\rm b}T}{3\eta_{\rm a}}\Biggl\{1 + \Gamma\left(\frac{4}{3}\right)\Gamma\left(\frac{2}{3}\right) \\ 
    + \frac{\lambda_{\rm a}}{r_{\rm c}}A\left[\Gamma\left(\frac{2}{3}\right) + \Gamma\left(\frac{4}{3}\right)\Gamma\left(\frac{1}{3}\right)\right]\Biggr\}N_{\rm c}^{2}.
\end{multline}
It is more difficult to ascertain the exact difference between the monodisperse rate and this formulation. 
However, assuming Kn $\approx$ 1, the rate is increased by a factor of $\approx$1.78 from the monodisperse formulation.

\subsection{Kn $\gg$ 1 regime}

For Brownian coagulation in the Kn $\gg$ 1 regime, the collisional kernel is given by \citep{Jacobson2005}
\begin{equation}
    K(r,r') = \pi\left(r + r'\right)^{2}\sqrt{\left(v(m)^{2} + v(m')^{2}\right)},
\end{equation}
where $v$ is the thermal velocity of the particle given by
\begin{equation}
    v = \sqrt{\frac{8k_{\rm b}T}{\pi m}}.
\end{equation}
Converting the kernel to mass units gives
\begin{equation}
\label{eq:k_gg1}
    K(m,m') = K_{0}\left(m^{1/3} + m'^{1/3}\right)^{2}\left(m^{-1} + m'^{-1}\right)^{1/2},
\end{equation}
where $K_{0}$ is the kinetic pre-factor
\begin{equation}
    K_{0} = \left(\frac{3}{4\pi\rho_{\rm d}}\right)^{2/3}\sqrt{8\pi k_{\rm b}T}.
\end{equation}
To make the integration analytically tractable later, we approximate the square root term as separable
\begin{equation}
    \left(m^{-1} + m'^{-1}\right)^{1/2} \approx H\left(m^{-1/2} + m'^{-1/2}\right),
\end{equation}
where $H$ is a fitting factor.
$H$ should be in the range [1/$\sqrt{2}$, 1], where 1/$\sqrt{2}$ corresponds to the monodisperse case and 1 a highly broadened size distribution.
We use a Brent–Dekker root finding method to find the value of $H$ that minimises the relative error across a typical Kn $\gg$ 1 mass range of 10$^{-20}$-10$^{-14}$ g ($r$ $\sim$ 1nm - 0.1$\mu$m),  yielding $H$ $\approx$ 0.85, which is approximately the mid value between 1/$\sqrt{2}$ and 1.

\subsubsection{Monodisperse solution}

For the monodisperse solution, the mass distribution takes the form $f(m)$ $\approx$ $N_{\rm c}$$\delta(m - \lambda)$ and  $f(m')$ $\approx$ $N_{\rm c}$$\delta(m' - \lambda)$.
The kernel now becomes
\begin{equation}
    K(m,m') = 4\sqrt{\frac{2}{\lambda}}K_{0}\lambda^{2/3}.
\end{equation}
Leading to a zeroth moment expression as
\begin{equation}
    \frac{dM^{(0)}}{dt} = -2\sqrt{\frac{2}{\lambda}}K_{0}\lambda^{2/3}.
\end{equation}
The full equation for the coagulation rate is then
\begin{equation}
    \left(\frac{dN_{\rm c}}{dt}\right)_{\rm coag} \approx -2\sqrt{\frac{2}{m_{\rm c}}}\left(\frac{3}{4\pi\rho_{\rm d}}\right)^{2/3}\sqrt{8\pi k_{\rm b}T}m_{\rm c}^{2/3}N_{\rm c}^{2},
\end{equation}
and simplifies to
\begin{equation}
    \left(\frac{dN_{\rm c}}{dt}\right)_{\rm coag} \approx - 8\sqrt{\frac{\pi k_{\rm b} T}{m_{\rm c}}}r_{\rm c}^{2}N_{\rm c}^{2},
\end{equation}
which is the same expression found in previous monodisperse studies \citep[e.g.][]{Lee2025}.

\subsubsection{Exponential distribution solution}

Finding a fully consistent analytical solution assuming an exponential distribution is difficult in this regime, but we can follow the same method in \citet{Moran2023} to derive a population averaged kernel function.
Placing Eq. \ref{eq:k_gg1} into Eq. \ref{eq:pop_av_k} gives
\begin{multline}
  \overline{K(m,m')} = \frac{1}{N_{\rm c}^{2}}\int_{0}^{\infty}\int_{0}^{\infty}K(m,m')f(m)f(m')dmdm' \\ 
  = \frac{K_{0}H}{N_{\rm c}^{2}}\int_{0}^{\infty}\int_{0}^{\infty}\left(m^{1/3} + m'^{1/3}\right)^{2}\left(m^{-1/2} + m'^{-1/2}\right) \\ \cdot f(m)f(m')dmdm'.
\end{multline}
This results in a set of six integrals, the solutions of which are non-integer powers of the moments. 
After integrating and some algebra, the end result is
\begin{multline}
    \overline{K(m,m')} = \frac{2K_{0}H}{N_{\rm c}^{2}} \\ \cdot\left(M^{(2/3)}M^{(-1/2)} + 2M^{(1/3)}M^{(-1/6)} + M^{(1/6)}M^{(0)}\right),
\end{multline}
where $M^{(2/3)}$, $M^{(-1/2)}$, $M^{(1/3)}$, $M^{(-1/6)}$ and $M^{(1/6)}$ are the $k$ = 2/3, $k$ = -1/2, $k$ = 1/3, $k$ = -1/6 and $k$ = 1/6 non-integer powers of the moments respectively. 
Using the moment generator for the exponential mass distribution (Eq. \ref{eq:mom_gen}), results in
\begin{equation}
\label{eq:K_av_l}
     \overline{K(m,m')} = 2K_{0}H\lambda^{1/6}\left[\Gamma\left(\frac{5}{3}\right)\Gamma\left(\frac{1}{2}\right) + 2\Gamma\left(\frac{4}{3}\right)\Gamma\left(\frac{5}{6}\right) + \Gamma\left(\frac{7}{6}\right)\right].
\end{equation}
Using the relation $\lambda^{1/6}$ = $\lambda^{2/3}$$\lambda^{-1/2}$, the final expression for the change in number density can be given in a similar form to the monodisperse relation
\begin{multline}
\label{eq:N_coag_exp_2}
    \left(\frac{dN_{\rm c}}{dt}\right)_{\rm coag} \approx -0.85\sqrt{\frac{8\pi k_{\rm b} T}{m_{\rm c}}}r_{\rm c}^{2} \\ \cdot\left[\Gamma\left(\frac{5}{3}\right)\Gamma\left(\frac{1}{2}\right) + 2\Gamma\left(\frac{4}{3}\right)\Gamma\left(\frac{5}{6}\right)  + \Gamma\left(\frac{7}{6}\right)\right]N_{\rm c}^{2}.
\end{multline}
 Overall, the exponential distribution collision rate is a factor of $\approx$1.37 faster than the monodisperse formulation in the Kn $\gg$ 1 regime.

\subsection{Collisional regime interpolation function}

Equations \ref{eq:N_coag_exp_1} and \ref{eq:N_coag_exp_2} are valid in the Kn $\ll$ 1 and Kn $\gg$ 1 limits respectively, however, in many cases particles occupy an intermediate regime Kn $\sim$ 1 where it is difficult to derive an analytical Brownian coagulation kernel. 
To solve this problem, many studies apply an interpolation function between the limiting regimes, with the most common following the \citet{Fuchs1964} prescription \citep[e.g.][]{Lavvas2011, Gao2018}.

In this study, we follow \citet{Moran2022} who propose a simple and accurate interpolation function between the continuum/diffusive (Kn $\ll$ 1) and free molecular/kinetic (Kn $\gg$ 1).
This is designed as a modification to the diffusive regime kernel as
\begin{equation}
    K(r,r') = 4\pi\left[D(r)  + D(r')\right](r + r') g({\rm Kn_{\rm D}}),
\end{equation}
where Kn$_{\rm D}$ is the diffusive particle Knudsen number given by
\begin{equation}
    {\rm Kn_{\rm D}} = \frac{8\sqrt{2}\left[D(r)  + D(r')\right]}{\pi\sqrt{v(r)^{2} + v(r')^{2}}\left(r + r'\right)},
\end{equation}
for a particle size pair.
The interpolation function, $g({\rm Kn_{\rm D}})$, derived in \citet{Moran2022} is stated as
\begin{equation}
    g({\rm Kn_{\rm D}}) = \left(1 + \frac{\pi^{2}}{8} {\rm Kn_{D}}^{2}\right)^{-1/2},
\end{equation}
which varies between a value of 1 for the diffusive/continuum regime as ${\rm Kn_{\rm D}} \rightarrow$ 0 and recovers the kinetic/free molecular kernel as ${\rm Kn_{\rm D}} \rightarrow$ $\infty$.
We can calculate a population averaged diffusive particle Knudsen number $\overline{{\rm Kn_{\rm D}}}$ using the population averaged kernels
\begin{equation}
    \overline{{\rm Kn_{\rm D}}} = \frac{2\sqrt{2}}{\pi} \frac{\overline{K_{\rm l}(m,m')}}{\overline{K_{\rm h}(m,m')}},
\end{equation}
where $\overline{K_{\rm l}(m,m')}$ is the population averaged kernel for the Kn $\ll$ 1 regime from Eq. \ref{eq:K_av_h2} and $\overline{K_{\rm h}(m,m')}$ is the population averaged kernel for the Kn $\gg$ 1 regime from Eq. \ref{eq:K_av_l}.

For a monodisperse size distribution, this interpolation function can be readily used to modify Eq. \ref{eq:N_coag1}
\begin{equation}
\label{eq:N_coag_1_int}
    \left(\frac{dN_{\rm c}}{dt}\right)_{\rm coag} \approx - \frac{4k_{\rm b}T\overline{\beta}}{3\eta_{\rm a}}N_{\rm c}^{2}g(\rm{Kn_{D}}),
\end{equation}
where $\rm{Kn_{D}}$ is evaluated at the mean grain size.
For the exponential distribution, the overall 8coagulation rate is given by
\begin{multline}
\label{eq:N_coag_exp_int}
    \left(\frac{dN_{\rm c}}{dt}\right)_{\rm coag} \approx - \frac{2k_{\rm b}T}{3\eta_{\rm a}}\Biggl\{1 + \Gamma\left(\frac{2}{3}\right)\Gamma\left(\frac{4}{3}\right) \\ 
    + \frac{\lambda_{\rm a}}{r_{\rm c}}A\left[\Gamma\left(\frac{2}{3}\right) + \Gamma\left(\frac{4}{3}\right)\Gamma\left(\frac{1}{3}\right)\right]\Biggr\}N_{\rm c}^{2}g\left({\rm \overline{Kn_{D}}}\right),
\end{multline}
which recovers the Kn $\gg$ 1 and Kn $\ll$ 1 collisional rate limits for the exponential distribution.

\subsection{Gravitational coalescence}

For gravitational coalescence, the collisional growth kernel is given by \citep[e.g.][]{Jacobson2005}
\begin{equation}
   \label{eq:grav_kern}
    K(r,r') = \pi(r + r')^{2}|v_{\rm f}(r) - v_{\rm f}(r')|E,
\end{equation}
where $v_{\rm f}$ [cm s$^{-1}$] is the settling velocity of the particle.
Defining $\Delta v_{\rm f} = |v_{\rm f}(r) - v_{\rm f}(r')|$ we follow \citet{Ohno2017} and evaluate $\Delta v_{\rm f}$ with a parameter, $\epsilon$, that estimates the relative velocity of the particles from the mean particle size settling velocity, $v_{\rm f}(r_{\rm c})$, giving $\Delta v_{\rm f}$ $\approx$ $\epsilon$$v_{\rm f}(r_{\rm c})$. 
This is taken as $\epsilon$ $\approx$ 0.5 following the results of \citet{Sato2016}, who found this value to best reproduce the results of a collisional bin resolving model for grain growth in protoplanetary disks.
$E$ is a collisional efficiency factor, dependent on the particle Stokes number.

\subsubsection{Monodisperse solution}

For the monodisperse distribution, putting the kernel from Eq. \ref{eq:grav_kern} into Eq. \ref{eq:N_coll} and evaluating at the mean particle size, leads to the expression found in previous studies \citep[e.g.][]{Lee2025}
\begin{equation}
  \left(\frac{dN_{\rm c}}{dt}\right)_{\rm coal} \approx -2\pi r_{\rm c}^{2}\Delta v_{\rm f}  E N_{\rm c}^{2}.
\end{equation}
In this case, E is given by the monodisperse properties with the Stokes number, Stk, given by
\begin{equation}
    {\rm Stk} = \frac{\Delta v_{\rm f} v_{\rm f}(r_{\rm c})}{gr_{\rm c}} = \frac{\epsilon v_{\rm f}^{2}(r_{\rm c})}{gr_{\rm c}}.
\end{equation}
$E$ is then \citep{Guillot2014}
\begin{equation}
     E =
    \begin{cases}
      {\rm max}\left[0, 1 - 0.42{\rm Stk}^{-0.75}\right]. & \overline{{\rm Kn}} < 1 \\
      1. &  \overline{{\rm Kn}} \geq 1 
    \end{cases}
\end{equation}

\subsubsection{Exponential distribution}

For an exponential distribution, we make the assumption that $\Delta$$v_{\rm f}$ and $E$ are well defined at the mean particle size value, allowing them to be taken outside the integral, making the integral analytically more tractable.
Defining the kinetic pre-factor as 
\begin{equation}
    K_{0} = \left(\frac{3}{4\pi\rho_{\rm d}}\right)^{2/3}\pi\Delta v_{\rm f} \overline{E},
\end{equation}
the population averaged kernel is expressed as
\begin{multline}
  \overline{K(m,m')} = \frac{1}{N_{\rm c}^{2}}\int_{0}^{\infty}\int_{0}^{\infty}K(m,m')f(m)f(m')dmdm' \\ 
  = \frac{K_{0}}{N_{\rm c}^{2}}\int_{0}^{\infty}\int_{0}^{\infty}\left(m^{1/3} + m'^{1/3}\right)^{2}f(m)f(m')dmdm'.
\end{multline}
Solving the integral results in
\begin{equation}
      \overline{K(m,m')} = \frac{2K_{0}}{N_{\rm c}^{2}}\left(M^{(2/3)}M^{(0)} + M^{(1/3)}M^{(1/3)}\right).
\end{equation}
In the same procedure as the coagulation kernels, assuming an exponential distribution results in the final expression being
\begin{equation}
      \left(\frac{dN_{\rm c}}{dt}\right)_{\rm coal} \approx -\pi r_{\rm c}^{2}\Delta v_{\rm f}\overline{E}\left[\Gamma\left(\frac{5}{3}\right) + \Gamma\left(\frac{4}{3}\right)^{2}\right]N_{\rm c}^{2}.
\end{equation}

For the polydisperse case, the population averaged Stokes number, $\overline{{\rm Stk}}$, is approximated using the number weighted mean radius, $\langle r\rangle$ [cm], rather than the mass weighted mean radius, $r_{\rm c}$,
\begin{equation}
    \overline{{\rm Stk}} = \frac{\Delta v_{\rm f} v_{\rm f}(r_{\rm c})}{g\langle r\rangle} = \frac{\epsilon v_{\rm f}^{2}(r_{\rm c})}{g\langle r\rangle},
\end{equation}
where the relation between the number and mass weighted radii is given by
\begin{equation}
    \langle r\rangle = r_{\rm c}\Gamma\left(\frac{4}{3}\right).
\end{equation}
$\overline{E}$ is then delineated using the number density averaged Knudsen number
\begin{equation}
     \overline{E} =
    \begin{cases}
      {\rm max}\left[0, 1 - 0.42\overline{{\rm Stk}}^{-0.75}\right]. & \overline{{\rm Kn}_{N}} < 1 \\
      1. &  \overline{{\rm Kn}_{N}} \geq 1 
    \end{cases}
\end{equation}
Overall, assuming the collisional efficiency factors are equal, the collisional growth rate from gravitational coalescence assuming an exponential distribution is $\approx$0.85 times slower than the monodisperse approximation.

\subsection{Summary of collisional growth results}

Similar to the condensation factor of 0.9, the $>$1.1 enhancement factor for Brownian motion in the Kn $\ll$ 1 regime makes intuitive sense for the exponential distribution as the distribution is more skewed towards smaller particles with an exponential drop-off for particle larger than the mean particle size.
The collision rate is therefore enhanced due to the consideration of the smaller particle population in the exponential distribution.
For the Kn $\gg$ 1 regime, the increase by $\approx$1.37 also makes intuitive sense, as in this regime collisions are guided by the relative cross-section of the particles which is reduced for smaller particles but more strongly counteracted by the increased thermal velocity of small particles.
Similarly, for gravitational coalescence, the collisional rate is dependent on the relative cross-section of the particles, resulting in an overall reduction in collisional growth rate of around $\approx$0.85 for the exponential distribution compared to the monodisperse distribution.

\section{Settling velocity interpolation function}
\label{sec:set_int}

\begin{figure}
    \centering
    \includegraphics[width=0.99\linewidth]{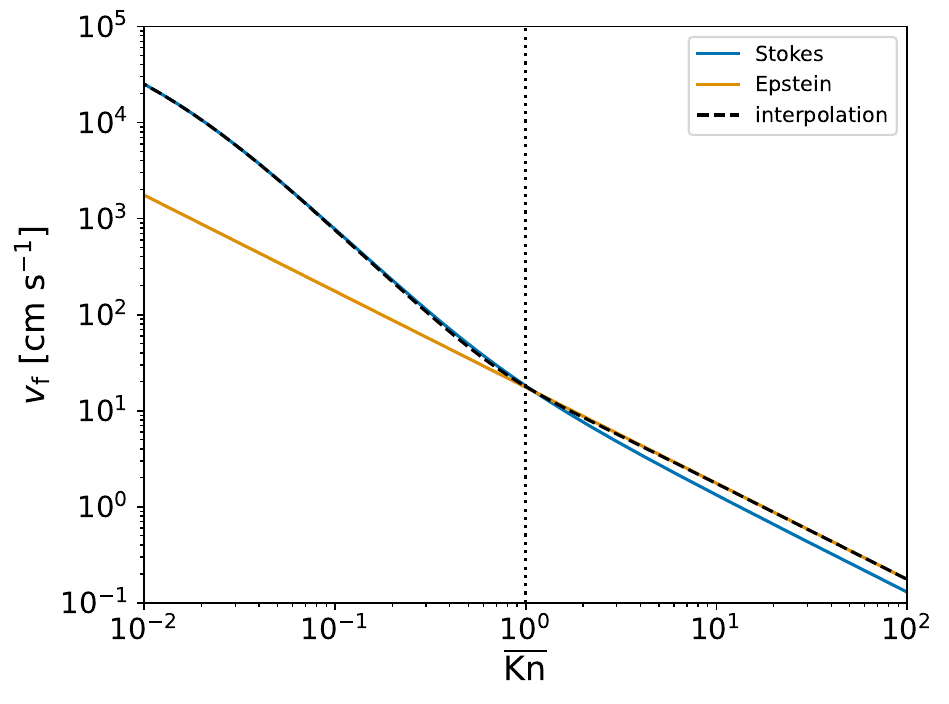}
    \caption{Example of the settling velocity tanh smooth interpolation scheme. 
    The blue line shows the Stokes law values (Eq. \ref{eq:vf_1}), the orange line the Epstein law values (Eq. \ref{eq:vf_2}) and the black dashed line the interpolation function (Eq. \ref{eq:vf_3}).
    The background gas is assumed to be T = 500 K, p = 10$^{-3}$ bar, log g = 3.25, $\overline{\mu}$ = 2.33 g mol$^{-1}$ and 100\% H$_{2}$.}
    \label{fig:vf}
\end{figure}

In the Kn $\ll$ 1 regime, the Stokes law with included Reynolds number dependent fitting factor can be used from \citet{Ohno2018}
\begin{equation}
\label{eq:vf_1}
    v_{\rm f}(r) = \frac{2\beta gr^{2}(\rho_{\rm d} - \rho_{\rm a})}{9\eta_{\rm a}}\left[1 + \left(\frac{0.45gr^{3}\rho_{\rm a}\rho_{\rm d}}{54\eta_{\rm a}^{2}}\right)^{2/5}\right]^{-5/4},
\end{equation}
where $g$ [cm s$^{-2}$] is the gravitational acceleration, and we have re-included the additional buoyancy term from the original Stokes law derivation.
However, in the Kn $\gg$ 1 regime, which is commonly a regime where small particles in exoplanet atmospheres are found, the Epstein regime drag law should be used.
Here, we can follow \citet{Woitke2003} who used the \citet{Schaaf1963} drag coefficients in the limiting case for settling velocities well below the atmospheric thermal velocity, $v_{\rm f}$ $\ll$ c$_{\rm T}$, to derive
\begin{equation}
\label{eq:vf_2}
    v_{\rm f}(r) = \frac{\sqrt{\pi} g \rho_{\rm d} r}{2\rho_{\rm a} c_{\rm T}},
\end{equation}
where $c_{\rm T}$ = $\sqrt{2k_{\rm b}T/m_{\rm a}}$ [cm s$^{-1}$] is the thermal velocity of the atmosphere.

We can again use the same tanh function from Eq. \ref{eq:tanh} to smoothly interpolate between the two expressions
\begin{equation}
\label{eq:vf_3}
    v_{\rm f} =  f({\rm Kn}')v_{\rm f, Stokes} + \left[1 - f({\rm Kn}')\right]v_{\rm f, Epstein},
\end{equation}
where $\rm{Kn}'$ = Kn/Kn$_{\rm cr}$.
Choosing a suitable critical value for the settling velocity transition region is a tricky endeavor, for example, 
\citet{Woitke2003} find a critical value of Kn$_{\rm cr}$ = 1/3 equating limits of their drag force derivations.
Through trial and error and visual inspection, we find that Kn$_{\rm cr}$ = 1 with a scaling factor of $a$ = 2 gives a good balance and smooth transition for the Knudsen number regime limit velocities.
Figure \ref{fig:vf} shows an example of this interpolation scheme for various Knudsen numbers for typical atmospheric properties.

\section{Y-dwarf KCl cloud simple 1D example}
\label{sec:1D}

\begin{figure*}
    \centering
    \includegraphics[width=0.48\linewidth]{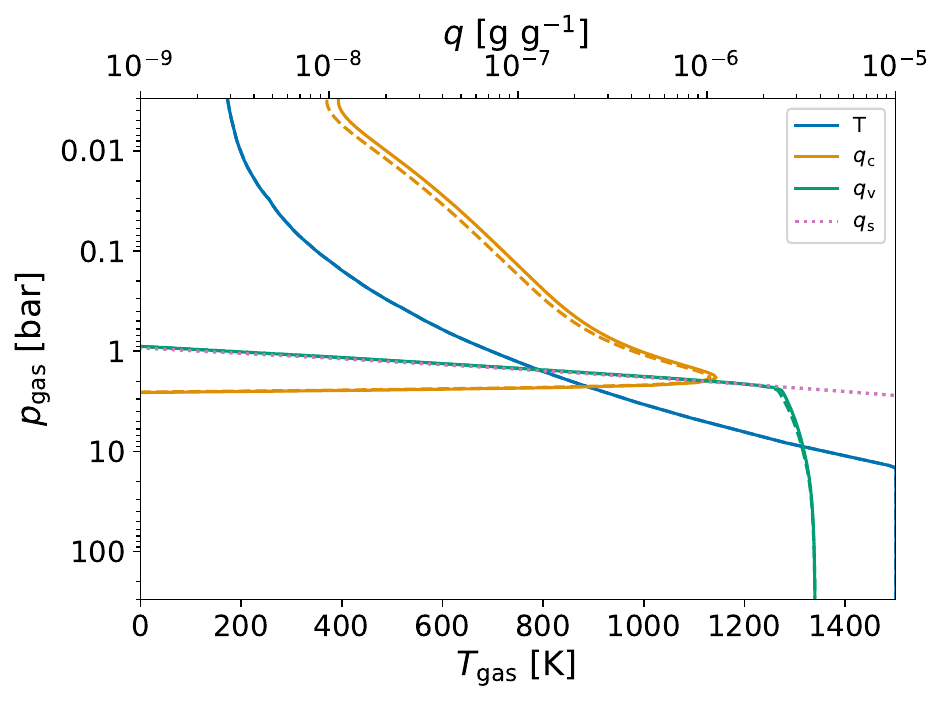}
    \includegraphics[width=0.48\linewidth]{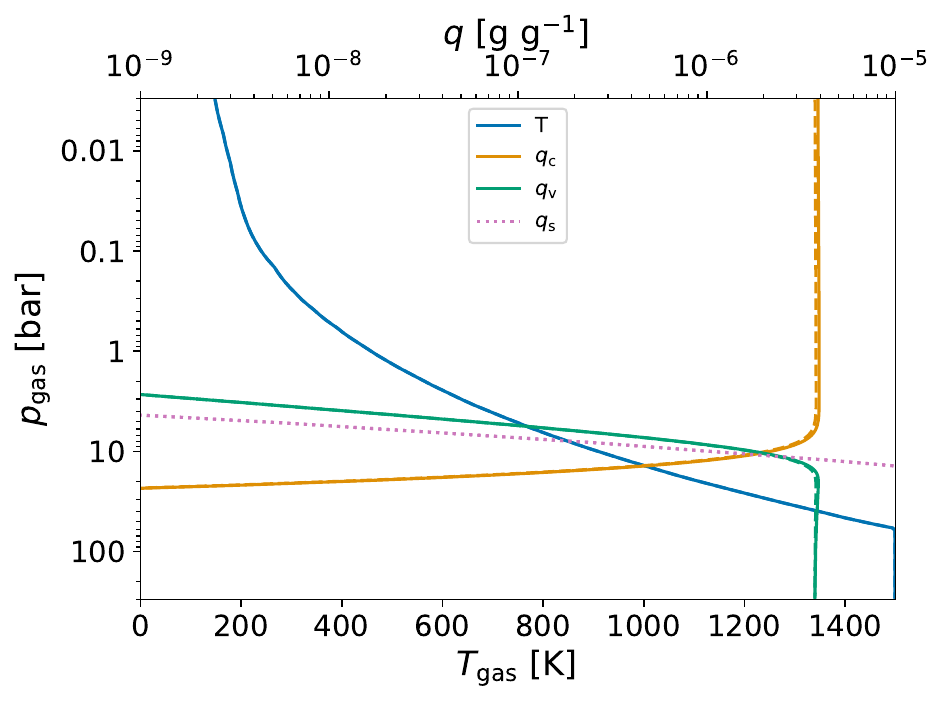}
    \includegraphics[width=0.48\linewidth]{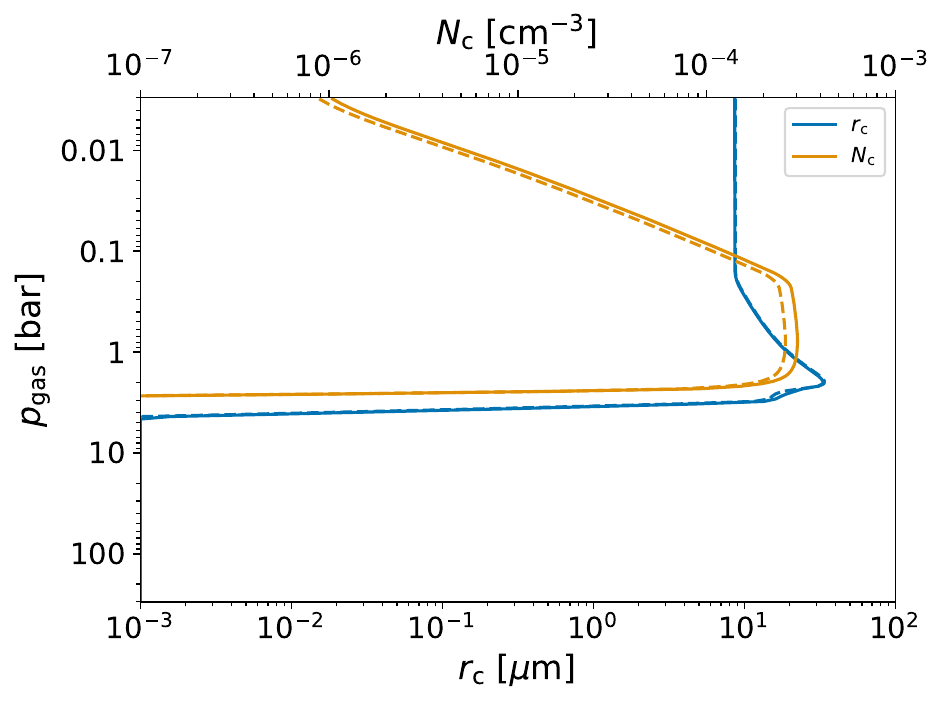}
    \includegraphics[width=0.48\linewidth]{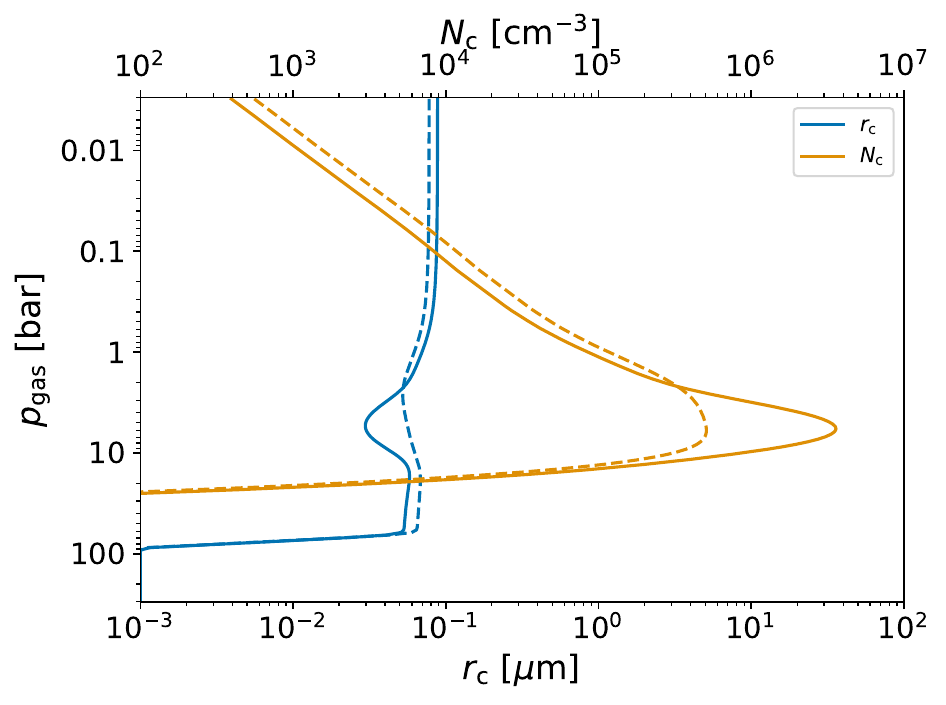}
    \includegraphics[width=0.48\linewidth]{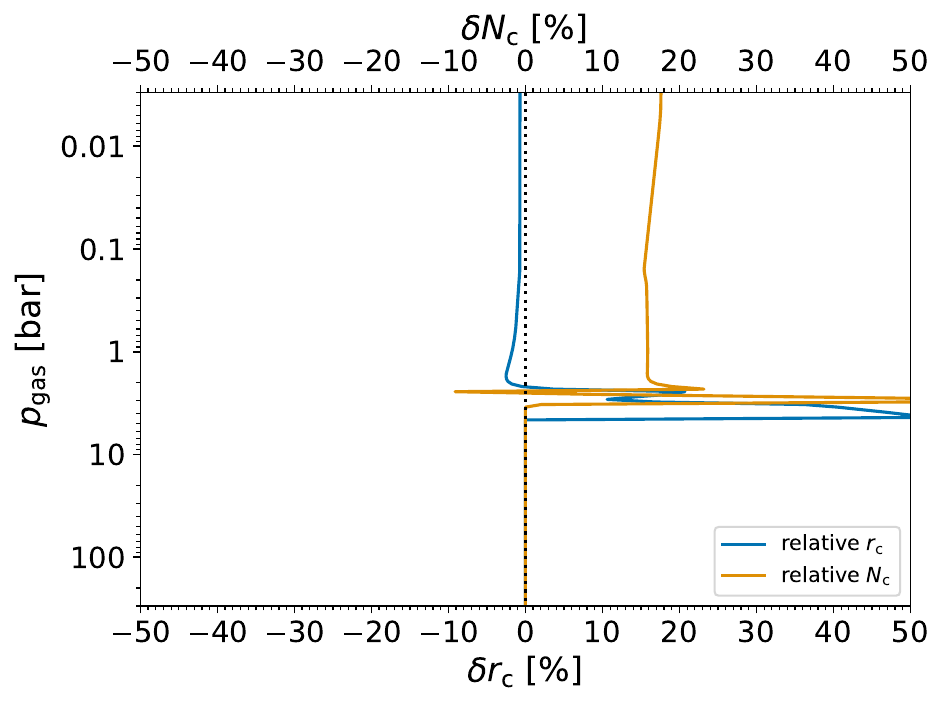}
    \includegraphics[width=0.48\linewidth]{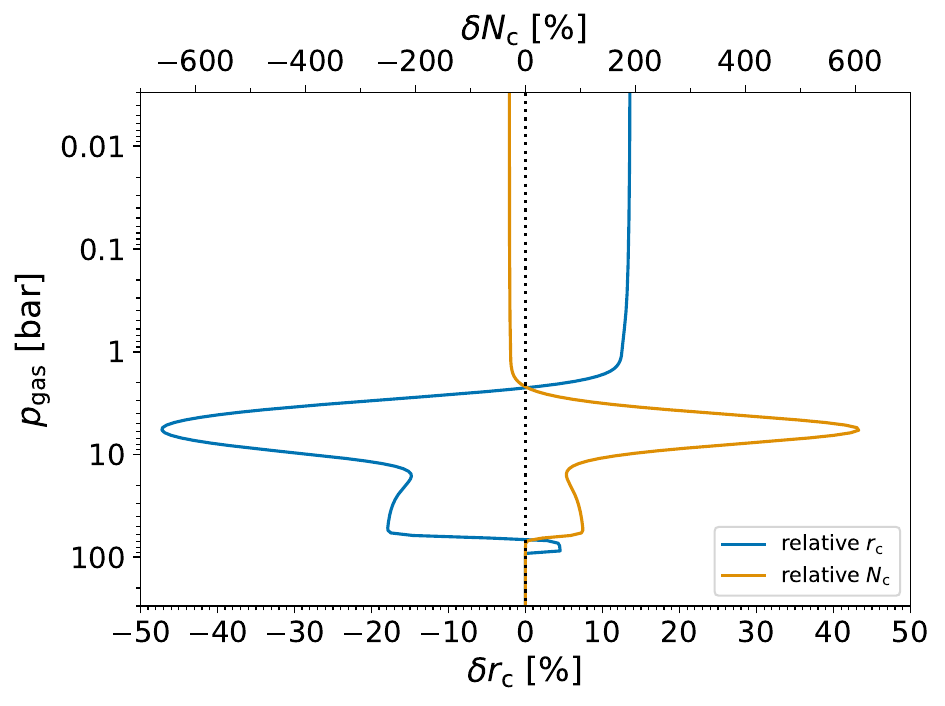}
    \caption{KCl cloud structures using the log g = 3.25 (left) and log g =  4.25 (right) Y-dwarf temperature-pressure profiles from \citet{Gao2018} and assuming a constant $K_{\rm zz}$ = 10$^{8}$ cm$^{2}$ s$^{-1}$.
    The dashed lines show the monodisperse size distribution and the solid lines the exponential size distribution results.
    The top panel shows the temperature-pressure profiles with the mass mixing ratio of the condensate, $q_{\rm c}$, vapour, $q_{\rm v}$ and saturation point, $q_{\rm s}$.
    The middle panel shows the mean particle size, $r_{\rm c}$, and total number density, $N_{\rm c}$. 
    The bottom panel shows the relative difference in $r_{\rm c}$ and $N_{\rm c}$ between the monodisperse and exponential cloud structure.}
    \label{fig:1D_KCl}
\end{figure*}

In this Section, we apply our new equation set to a simple 1D KCl Y-dwarf example.
We follow the same setup from \citet{Gao2018}, taking their Y-dwarf temperature-pressure (T-p) profiles, log g and constant $K_{\rm zz}$ parameters used in their study.
We use a simple, second order explicit 1D vertical diffusion and advection time stepping scheme, similar to that used in recent Exo-FMS GCM simulations of brown dwarfs \citep{Lee2024, Lee2025}.
For the cloud microphysics, we use the ODE system presented in \citet{Lee2025} which includes condensation/evaporation, collisional growth and homogeneous nucleation, but with the modifications for the exponential distribution rates from this study when appropriate.
We assume that the settling velocity for each moment is well defined at the mass-weighted radius (Eq. \ref{eq:rc}) as derived from the moment solutions at each timestep.
We also assume that the particles are well coupled to the turbulent flow, and so are also vertically diffused in the atmosphere at the given $K_{\rm zz}$ eddy diffusion coefficient \citep[e.g][]{Woitke2020}.
We integrate the 1D column in time for 10$^{9}$ seconds, after which all variables show relative differences less than 10$^{-6}$ between timesteps.
We plot the relative difference between the exponential and monodisperse results as
\begin{equation}
    \delta N_{\rm c} = \frac{N_{\rm c}({\rm exp}) - N_{\rm c}({\rm mono})}{N_{\rm c}({\rm mono})},
\end{equation}
with the equivalent expression for the mean particle size.

Figure \ref{fig:1D_KCl} shows our results for the 1D column model. 
In the log g = 3.25 case, we find around a $\sim$15\% increase in total number density in the exponential distribution compared to the monodisperse distribution, with the mean particle sizes differing by $\sim$1-2\%.
We suggest that, due to the mean particle size being quite large in this case, $\sim$10 $\mu$m, this increase in number density arises from the decrease in the coalescence rate between the monodisperse and exponential distributions.
This is reflected in the slightly smaller particle sizes seen for the exponential case.

In the log g = 4.25 case, larger differences are seen between the size distribution assumptions, with around a 25\% reduction in cloud particle number density in the exponential case compared to the monodisperse case in the upper atmosphere, but a $>$200\% difference closer to the cloud base. 
A $\sim$15\% relative difference is seen in mean particle radius in the upper atmosphere, but this increases to $>$40\% at the cloud base.
Due to the small mean particle size of $\sim$0.1 $\mu$m, this suggests that the increase in coagulation rate in the Kn $\gg$ 1 regime is responsible for this difference in the upper atmosphere, while slower condensation occurring near the cloud base accounts for the increased number density and decreased particle radii.

In both size distribution cases, the mass mixing ratio and vapour mixing ratio profile remain highly similar.
This suggests that the total mass that is condensed is not markedly altered between each assumption, but the balance in mass distributed between number density and particle size is changed.

\section{Discussion}
\label{sec:disc}

\begin{figure*}
    \centering
    \includegraphics[width=0.49\linewidth]{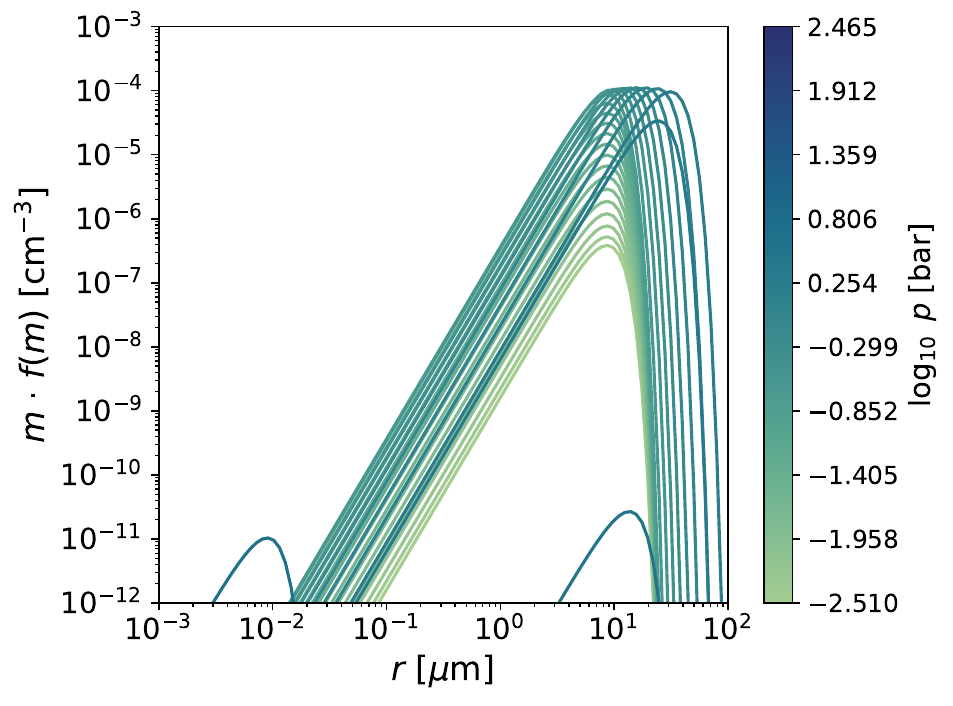}
    \includegraphics[width=0.49\linewidth]{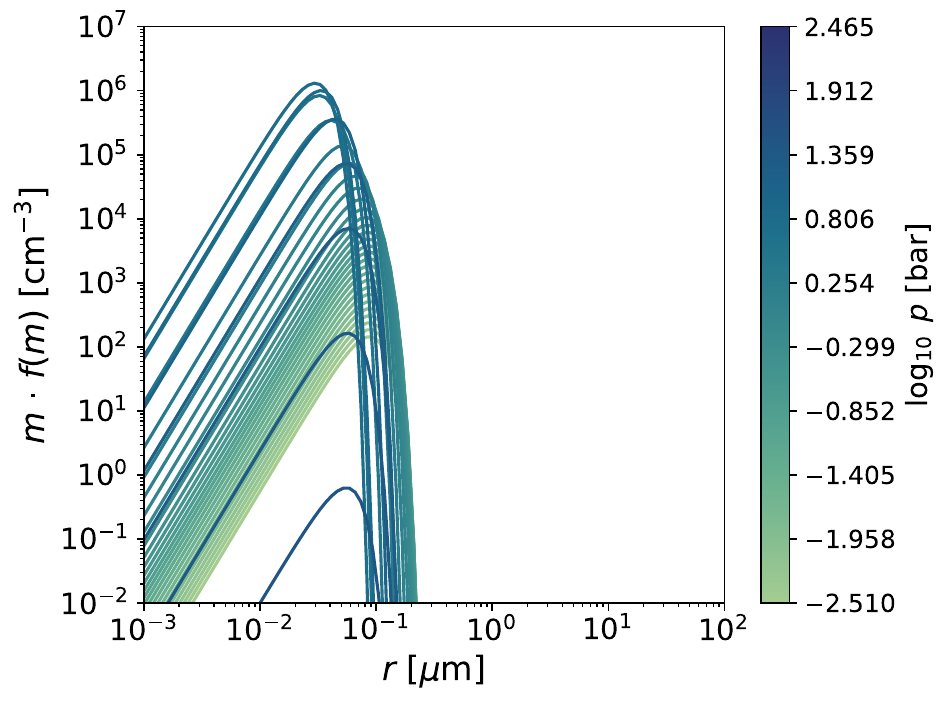}  
    \caption{Reconstructed exponential particle size distributions (Eq. \ref{eq:exp_dist}) from the Y-dwarf KCl case moment results (Left: log g = 3.25, Right: log g = 4.25) in Section \ref{sec:1D} when the exponential distribution is assumed. 
    The colour bar shows the atmospheric pressure of the distribution.
    This shows the limitations of the exponential distribution as limited flexibility is present due to the assumed variance, resulting in a highly similar distribution shape across the cloud layer. 
    }
    \label{fig:exp_recon}
\end{figure*}

Several assumptions were made to simplify the coagulation collisional growth integrals analytically.
Specifically, in the Kn $\ll$ 1 regime we assumed a linear Cunningham slip factor which has around a 10\% error near the Kn $\sim$ 1 transition regime.
Possibilities to improve accuracy in this regime would be to fit a quadratic equation to the Cunningham slip $\beta$ function which would still enable an analytic solution, but increase the number of gamma correction terms in the Kn $\ll$ 1 coagulation rate expression.
In the Kn $\gg$ 1 regime, beyond fitting for the $H$ factor, no further assumptions were required to produce a solution.
For gravitational coalescence, we assumed the collisional efficiency factor was well defined at the mean particle size.
However, this assumption may not be entirely accurate, particularly in the Kn $<$ 1 regime, where efficiency factors can vary widely, including reaching zero.
Should the efficiency factor vary significantly from the average value across the size distribution, we can expect a significantly altered overall coalescence rate to those presented here.
Accounting for these additional effects analytically or through a numerical or interpolation approach and relaxing the assumptions may be possible in the future, but is beyond the scope of the current study.

Our assumption of an exponential size distribution has many limitations, especially when compared to more flexible size distributions such as the gamma or log-normal distribution.
The single parameter limitation of the exponential distribution inherits the assumption that the overall particle size distribution shape does not change in time, with the variance remaining constant at the square of the average particle mass at across the time evolution of the cloud structure.
The gamma and log-normal distribution contain an additional parameter compared to the one-parameter exponential distribution, allowing them to capture a much less strict and flexible size distribution shape.
A major advantage of the more complex distributions would be the allowance of the representation of both narrow and broad size distributions through directly calculating the variance using the moment results, rather than the assumed variance of the exponential distribution.
To visualise the exponential distribution limitations, in Figure \ref{fig:exp_recon} we show the reconstructed exponential distributions for the results in Section \ref{sec:1D} that used the exponential equation set from this study.
From this, it is clear that the overall shape of the distribution does not change much across the cloud structure, a direct result of the assumed variance in the exponential distribution, meaning representation of a narrower or broader size distribution is limited in the exponential distribution.
This can be compared to when the variance can be flexible, for example, \citet{Helling2008} reconstruct a potential exponential distribution that produced both narrow and broad distributions with height.
However, it should be noted that \citet{Helling2008} reconstructed the size distribution from a monodisperse microphysics method.

However, with its useful analytical properties and single parameter enabling a two-moment representation, rather than three or more moments for the more complex distributions, the exponential distribution offers a simple way to attempt to take into account polydispersity in the size distribution.
With the equations presented here, modellers can be more self-consistent in exploring the effects of polydispersity, rather than reconstructing distributions from moment solutions obtained from the monodisperse equation set \citep[e.g.][]{Helling2008, Stark2015, Lee2023}.
Since the exponential distribution is a special parameterised form of the gamma distribution, we suspect that designing a three-moment framework with the gamma distribution will take a similar form to that derived here but with additional correction terms to the monodisperse equation set.

Our results in Section \ref{sec:1D} used the same Y-dwarf T-p profiles as \citet{Gao2018}, which also allows us to compare the moment method results to the 1D CARMA bin-resolving model used in \citet{Gao2018} who also modelled pure KCl cloud particle formation.
Despite the exact detailed condensation, nucleation and collisional growth equations being different between this study and those in \citet{Gao2018}, this serves as a useful comparison between both methodologies.
From the results in \citet{Gao2018} in the log g = 3.25 case the results between the presented moment method and CARMA are remarkably similar.
Both models show a similar condensed mass fraction structure.
The mean particle size also shows good agreement in this case, hovering around the 10 $\mu$m range, but the CARMA results tend to produce slightly smaller particles in the upper atmosphere by a few factors in this case.
However, it is difficult to one-to-one compare, since \citet{Gao2018} present the effective particle radius (weighted by area) whereas our results show the mass-weighted radius (weighted by mass/volume). 
Despite this, our results are quite consistent as the effective particle radius is always smaller than the mass-weighted radius by some factor in a particle size distribution.
In the log g = 4.25 case, our results are much more different to \citet{Gao2018}, in particular \citet{Gao2018} shows a near constant effective particle radius of around 3 $\mu$m whereas we produce particles at around $\sim$0.08 $\mu$m.
We suggest that this may be a case where the bin model produces a bimodal distribution, where the particle size distribution is split between a small particle regime and large particle regime, not able to be captured by the current moment methodology.
This skews the effective particle radius to the large particle population in the bin model, even if they are significantly less in number density.
We suspect that the moment method is capturing the small particle regime in this case but is unable to model the large particle population due to the assumption of a continuous distribution inherent in the method.

\section{Conclusions}
\label{sec:conc}

In this short paper, we derived condensation/evaporation and collisional growth expressions for a two moment cloud microphysical scheme applicable to sub-stellar atmospheric clouds.
We went beyond the monodisperse assumptions commonly used in previous studies and found simple analytical and tractable solutions when assuming an exponential particle size distribution.

Our analysis found that assuming an exponential distribution for particles leads to simple analytical formulae, which, in effect, modify the monodisperse formulations with constant factors for the time dependent moment equations.
These are $\approx$0.9 for the condensation/evaporation rate (first moment) and $>$1.1 for Brownian collisional growth (zeroth moment) in the Kn $\ll$ 1 regime and $\approx$1.37 for Brownian collisional growth in the Kn $\gg$ 1 regime and $\approx$0.85 for the gravitational coalescence collisional rate.
However, we note that several simplifying assumptions were made regarding integrating the collision growth equations in order to make the problem more analytically tractable. 
In addition, we proposed simple interpolation functions for evaluating the condensation, settling velocity and collisional growth rates in the Kn $\sim$ 1 transition regime.

In a simple 1D column model example, we found significant differences of a factor of $>$200\% in number density and $>$40\% in particle size between the monodisperse and exponential size distribution assumptions, with differences dependent on the T-p profile and gravity of the simulation.
We suggest that, in our particular cases, the total condensed mass in the cloud structure is similar between each assumption, but the balance between number density and particle sizes is greatly affected by the choice of size distribution assumption.

In future papers, we will extend our two moment framework to three moments to enable self-consistent evolution of more complex, three parameter, size distributions, but are still somewhat analytically tractable, such as the log-normal and gamma distributions.
Due to the exponential distribution case being a special case of the gamma distribution, we find it likely that derivations using the gamma distribution will follow similar structure to the equations presented in this study.

\begin{acknowledgements}
E.K.H. Lee is supported by the CSH through the Bernoulli Fellowship.
We thank K. Ohno for initial discussions on the topic.
We thank the referee for suggestions that led to improving the consistency of the two-moment framework.
\end{acknowledgements}

\bibliographystyle{aa} 
\bibliography{bib.bib} 


\end{document}